\documentstyle[11pt,newpasp,epsf,twoside]{article}
\def\edcomment#1{\iffalse\marginpar{\raggedright\sl#1\/}\else\relax\fi}
\reversemarginpar

\begin{document}
\title{{\bf FIRES} at the VLT: Measuring the Rest-Frame V-Band Luminosity of Galaxies from z$\sim$3 to Now}

\vspace{-0.5cm}

\author{Gregory Rudnick$^1$, Hans-Walter Rix$^1$, \& Marijn Franx$^2$}
\affil{$^1$MPIA, K\"onigstuhl 17, Heidelberg D-69117, Germany}
\affil{$^2$Leiden Observatory, PO BOX 9513, 2300 RA Leiden, Netherlands}

\vspace{-0.2cm}

\begin{abstract}
  
  We present early results from the \textbf{F}aint
  \textbf{I}nfra\textbf{R}ed \textbf{E}xtragalactic \textbf{S}urvey
  (FIRES) at the VLT, the main goal of which is to study galaxy
  evolution in a deep, K-band selected sample.  With our NIR selection,
  we select galaxies based on their rest-frame optical light at all
  redshifts $z\la3$.  Our seven band photometry, coupled with an
  accurate and reliable photometric redshift technique, gives us the
  ability to study galaxies at the same rest-frame wavelength across a
  large range in redshift.  We present here the rest-frame V-band
  luminosity $L_{V}$ of objects in our sample as a function of
  redshift and demonstrate the importance of near infrared data in
  measuring the correct $z_{phot}$ and $L_{V}$.
\end{abstract}


Pre-existing optical surveys, which have been very successful in
finding large spectroscopically confirmed populations of $z>2.5$
galaxies (e.g.  Steidel et al. 1996), select these galaxies by their
rest-frame far ultraviolet (UV) light.  This light is dominated by the
most massive stars in a galaxy and these searches may not sample the
population of galaxies which dominates the stellar mass at these
redshifts.  By looking in the near infrared (NIR), one can select
galaxies at high redshift by their rest-frame optical light.  The
rest-frame optical is less affected by dust extinction than the
rest-frame UV, and is also a better tracer of the older stars which
dominate the stellar mass of a galaxy (i.e. the luminosity weighted age
of a galaxy is older in the optical than in the UV).

FIRES is a program to image $\sim 30~arcmin^2$ of sky in $J_sHK_s$
with ISAAC at the VLT.  FIRES supplements deep HST optical images of
the Hubble Deep Field South (HDF-S) and the cluster MS1054-03 (van
Dokkum et al. 1998) with deep, high spatial resolution VLT data.  For
the HDF-S, we have fully reduced the first $\sim6$ hours (out of a
total of 32 hours) of exposure time in each passband.  The seeing in
the combined $K_s$-band image is $\approx0\farcs45$.

From the $K_s$-band image, we have generated a preliminary catalog
containing 345 objects using the SExtractor software (Bertin \&
Arnouts 1996).  Of these, 41 had sizes in the F814W WFPC2 image
consistent with being point sources.  In the $K_s$-band, we have
determined from simulations that we are $50\%$ complete for small
sources down to $K_s(AB)\approx24$.  Therefore, we further limited our
sample to those galaxies with magnitudes $K_s(AB)\leq 24$ (190
objects).  After convolving final images in each filter to the same
seeing, we measured the fluxes of the catalog objects in all the
bandpasses using a fixed 2\arcsec aperture.

We then measured the photometric redshift of each object using a
template fitting algorithm which employs the empirical galaxy spectral
templates of Coleman, Weedman, \& Wu (1980) adjusted for intergalactic
HI absorption (Madau 1995).  At every redshift we find the best linear
combination of templates which fit the observed SED and choose the
most likely redshift as that with the lowest chisquared.  When testing
our photometric redshift technique on objects with published
photometry and spectroscopic redshifts in the HDF-N, we found
$\sigma_z/(1+z_{spec})\approx 0.07$ across the entire redshift range.
We then measured the interpolated, rest-frame V-Band luminosity from
the best fit template combination at the best fit redshift (assuming
$\Omega_m=0.3, \Omega_{\Lambda}=0$).  

The importance of the NIR data points in measuring both the correct
redshift and the correct rest-frame V-Band luminosity is demonstrated
in Fig. 1.  In Fig. 2 we plot the rest-frame V-band luminosity vs.
redshift for the 190 objects with $K_s(AB)~\leq~24$ and $z~\leq~3$.  Note
the number of luminous galaxies at high redshift.

\begin{figure}
\plottwo{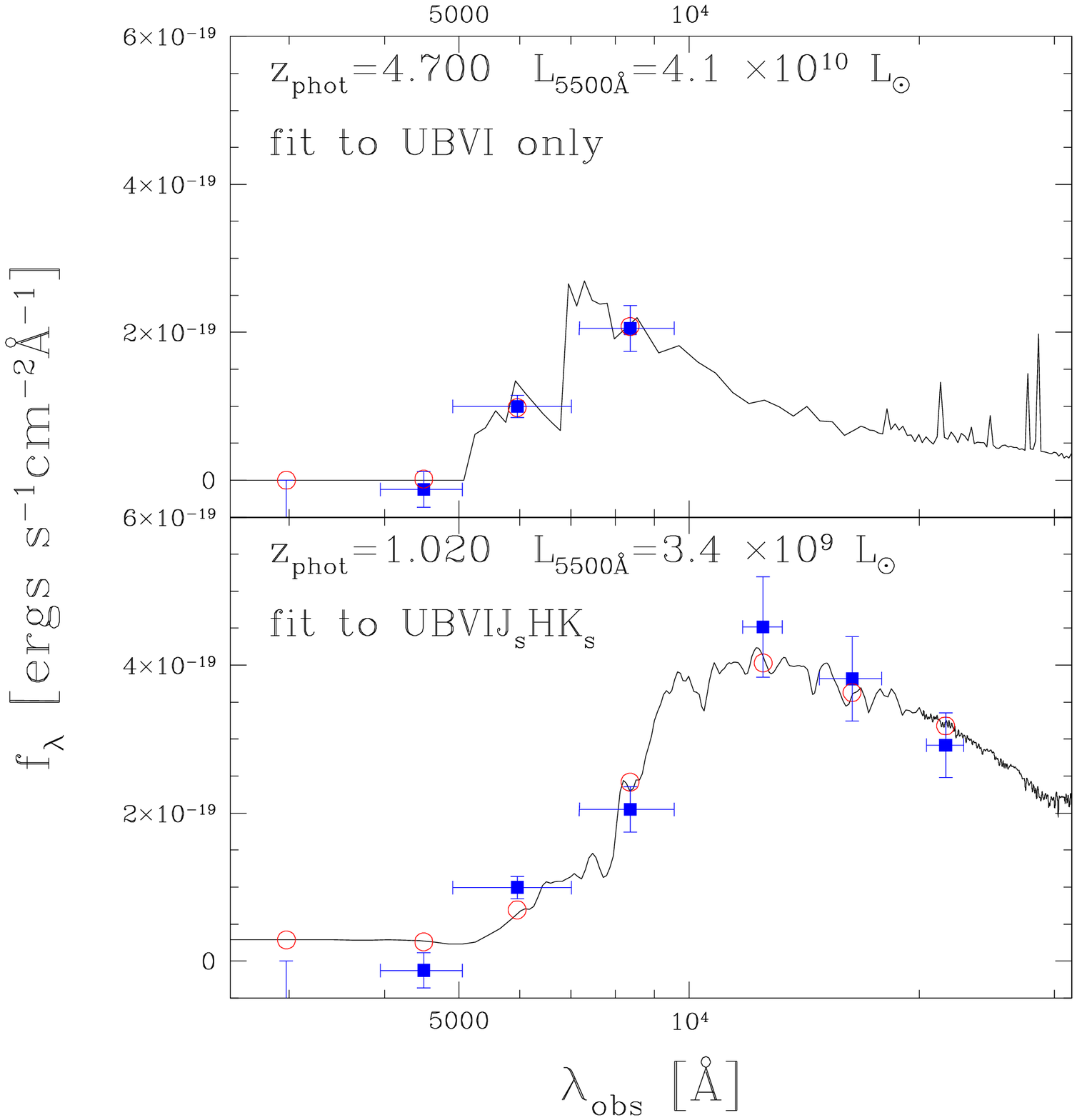}{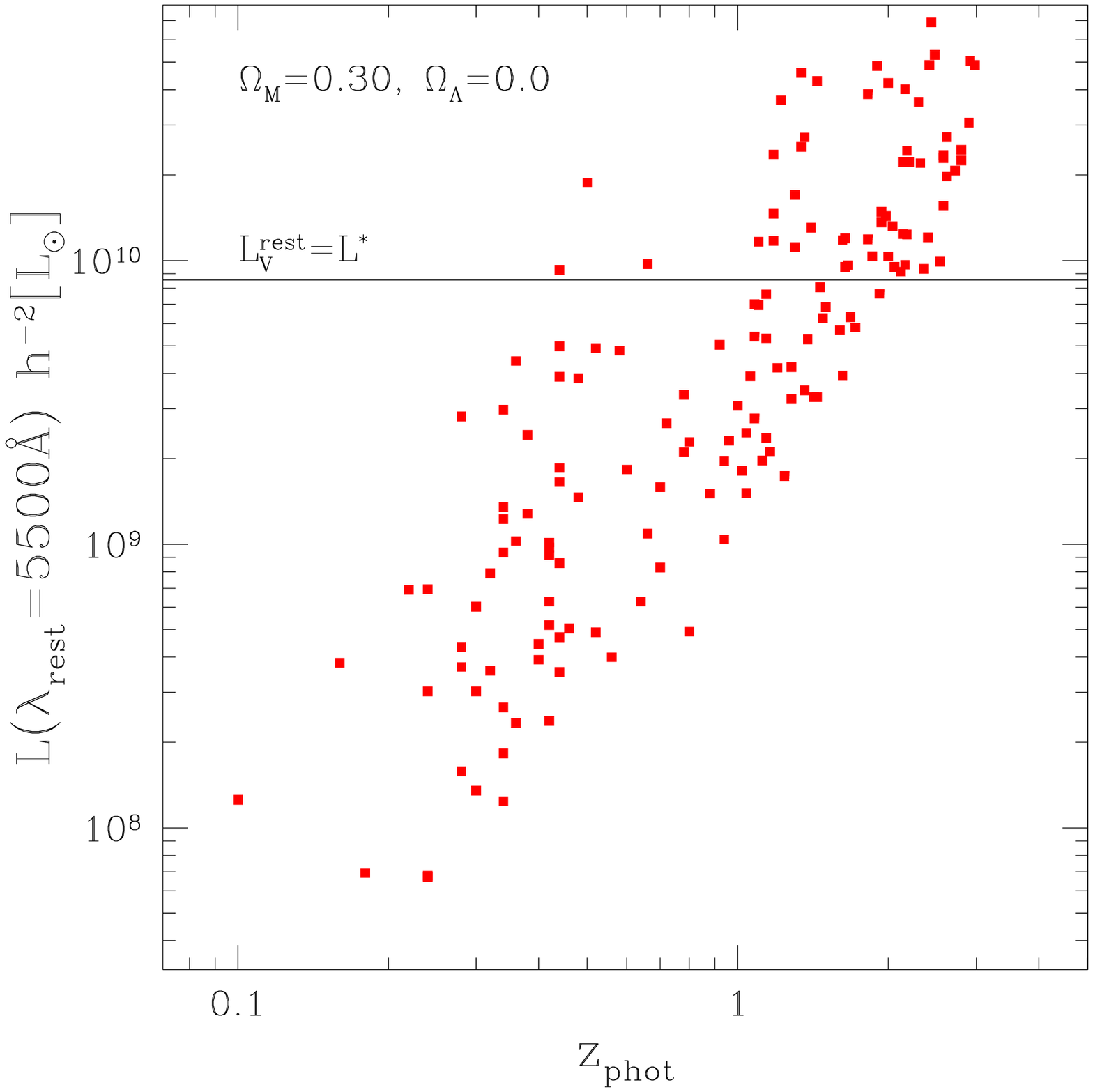}
\begin{flushleft}\parbox[t]{6.5cm} {
    {\bf Figure 1.}  We show, in both panels, the best fit
    template to the data at the best redshift $z_{phot}$.  The {\it
      top} panel shows the fit with only the optical HST data, while
    the {\it bottom} panel shows the fit with the ground-based NIR
    data included.  The solid squares with error bars are the actual
    data while the open circles are the model flux points.}
\end{flushleft}

\vspace{-4.5cm}
\begin{flushright}\parbox[t]{6.5cm} {
    {\bf Figure 2.} The distribution of rest-frame V-band
    luminosities as a function of photometric redshift.  We show all
    190 galaxies with $z_{phot}~\leq~3$ and $K_s(AB)~\leq~24$.  Note the
    large range in luminosities and the large number of intrinsically
    bright galaxies at $z_{phot}~\ga~1$.}
\end{flushright}
\vspace{-.1cm}
\end{figure}

\vspace{-0.5cm}
{}

\end{document}